\documentclass[12pt]{iopart}
\usepackage{graphicx}
\usepackage{multirow}
\usepackage{units}
\usepackage[usenames]{color}

\def\vk{{\bf k}}

\begin{document}

\title[]{Electronic structure of rare-earth mononitrides: quasiatomic excitations and semiconducting bands}

\author{Anna Galler$^1$ and Leonid V. Pourovskii$^{2,3}$}
\address{$^1$Institute of Solid State Physics, TU Wien, Wiedner Hauptstrasse 8-10, 1040 Vienna, Austria\\
$^2$Centre de Physique Th\'eorique, Ecole Polytechnique, CNRS, Institut Polytechnique de Paris, 91128 Palaiseau Cedex, France\\
$^3$Coll\`ege de France, 11 place Marcelin Berthelot, 75005 Paris, France
}
\ead{anna.galler@tuwien.ac.at}

\vspace{10pt}
\begin{indented}
\item[]\today
\end{indented}

\begin{abstract}
The electronic structure of the rare-earth mononitrides $Ln$N (where $Ln=$rare-earth), which are promising materials for future spintronics applications, is difficult to resolve experimentally due to a strong influence of defects on their transport and optical properties.  At the same time, $Ln$N  are challenging for theory, since wide semiconducting/semimetallic  2$p$ and 5$d$ bands need to be described simultaneously with strongly correlated 4$f$ states.
Here, we calculate the many-body spectral functions and optical gaps of a series of $Ln$N (with $Ln=$ Pr, Nd, Sm, Gd, Tb, Dy, Ho, Er)  by a density-functional+dynamical mean-field theory (DFT+DMFT) approach treating the correlated $Ln$ 4$f$ shells within the quasi-atomic Hubbard-I approximation. The on-site Coulomb interaction in the 4$f$ shell  is evaluated by a constrained DFT+Hubbard-I approach. Furthermore, to improve the treatment of semiconducting bands in DFT+DMFT, we employ the modified Becke-Johnson semilocal exchange potential.   Focusing on the paramagnetic high-temperature phase, we find that all investigated $Ln$N are  $pd$ semiconductors with gap values ranging from \unit[1.02]{} to \unit[2.14]{eV} along the series. The $pd$  band gap is direct for light $Ln=$ La...Sm and becomes indirect for heavy rare-earths. Despite a pronounced evolution of the $Ln$ 4$f$ states along the series,  empty 4$f$ states are invariably found above the bottom of the 5$d$ conduction band. The calculated spectra agree well with those available from x-ray photoemission, x-ray emission and x-ray absorption measurements. 
\end{abstract}

\section{Introduction}
The mononitrides $Ln$N (where $Ln$ is a lanthanide element) represent a rare case of ferromagnetic semicoductors/semimetals~\cite{Trodahl2007_GdN,Meyer2008,Lambrecht2013,Hewett2019_NdN,Hewett2020_DyN}   standing out among other lanthanide pnictides, which are generally antiferromagnets~\cite{Duan2007}. In the ferromagnetic state, both the top of the valence and bottom of the conduction band in $Ln$N  are expected to be of majority-spin character resulting in a complete spin polarization of hole and electron carriers~\cite{Lambrecht2013}.  The mononitride series is thus promising for spintronics applications~\cite{Senapati2011,Caruso2019}. It has attracted a renewed interest recently, especially from the experimental side, with detailed measurements of the optical conductivity in NdN~\cite{Hewett2019_NdN}, SmN~\cite{Hewett2019_SmN,Azeem2018_SmN} and DyN~\cite{Hewett2020_DyN}, quantitative studies of the effects of nitrogen vacancies in SmN~\cite{Azeem2019_SmN} and DyN~\cite{Hewett2020_DyN} and attempts to reconstruct the band structure of NdN~\cite{Hewett2019_NdN} and SmN~\cite{Hewett2019_SmN} from optical measurements. 

Even if the $Ln$N compounds have been   studied both experimentally and theoretically for several decades, there still exist numerous ambiguities regarding their electronic structure, transport and magnetic properties. Experimental investigations are often hindered by difficulties in fabricating good stoichiometric samples free of oxygen impurities and nitrogen vacancies. Each nitrogen vacancy  is predicted to dope two electrons to the conduction band with the third one forming a state in the gap~\cite{Punya2011}. Apparently, due to the effect of vacancies, earlier transport measurements on  $Ln$N samples often reported a metallic behavior~\cite{Wachter1980,Degiorgi1990,Wachter1998}.  
With progress in the fabrication of high-quality $Ln$N films by pulsed laser deposition and molecular beam epitaxy, the semiconducting nature of several $Ln$N has been established, including  GdN~\cite{Granville2006,Trodahl2007_GdN}, 
 NdN~\cite{Trodahl2016}, SmN~\cite{Preston2007} and DyN~\cite{Preston2007,Hewett2020_DyN}. 
Separating out the effect of defects remains, however, an outstanding issue. Vacancy-induced levels within the $pd$ gap  can be difficult to distinguish from sharp 4$f$ states resulting in conflicting reports on the nature of states forming the bottom of the conduction band in NdN, SmN, and DyN~\cite{Azeem2013_DyN,Azeem2018_SmN,Hewett2019_NdN,Hewett2019_SmN,Hewett2020_DyN}. In addition,  $Ln$N  are  prompt to rapid oxidation, hence, most of recent investigations are carried out on $Ln$N thin films grown on semiconducting substrates and protected by capping layers~\cite{Lambrecht2013}.  Intrinsic properties of bulk stoichiometric $Ln$N are thus hard to access experimentally.

In view of these difficulties, there have been numerous {\it ab initio} studies  aimed at establishing the electronic structure of pristine  $Ln$N.
First band structure calculations of Gd-pnictides were carried out already in the 70s~\cite{Hasegawa1977}.
However, DFT in conjunction with the standard local density approximation (LDA) or generalized-gradient approximation (GGA) treatments of exchange and correlation effects has a well-known tendency to underestimate band gaps in semiconductors.  
In addition,  $Ln$ 4$f$ electrons exhibit strong local correlation effects that are beyond standard DFT  approaches. 
Hence, a first overarching study of the electronic structure of the entire series of rare-earth mononitrides $Ln$N ($Ln=$ Ce...Yb)  by Aerts \textit{et al.}~\cite{Aerts2004} employed a self-interaction corrected (SIC) approach to take into account the localized nature of the $Ln$ 4$f$ states. This study predicted a broad range of electronic properties 	along the series, from half-metallic to insulating.   Another comprehensive study of the $Ln$N electronic structure was performed by Larson \textit{et al.}~\cite{Larson2007} employing DFT with $+U$ correction applied to $Ln$ 4$f$. 
Since a magnetic order needs to be assumed in DFT+$U$ to open the Mott gap, the paramagnetic electronic structure was derived by averaging over spin up and down contributions.  
With the Hubbard $U$ applied to the $Ln$ 4$f$ states only, DFT+$U$ predicted half-metallic states for the $Ln$N series~\cite{Larson2007}. 
In order to correct the underestimated semiconducting $pd$ gap,  Refs.~\cite{Larson2007,Morari2015} introduced a $+U$ term also for the empty 5$d$ shell, for which  it amounts to an upward shift of the 5$d$ states by the double-counting term. Though this shift can indeed correct the gap in an {\it ad hoc} way,  the underlying physics of {\it nonlocal} exchange opening the semiconducting band gap is not captured by this approach.

The electronic structure of ferromagnetic $Ln$N were subsequently calculated using more advanced approaches to nonlocal exchange---hybrid functionals and the G$W$ method.
Employing  the hybrid functional B3LYP, Ref.~\cite{Doll2008} predicted a half-metallic ground state for ferromagnetic GdN,  while another hybrid functional (HSE) calculation~\cite{Schlipf2011} obtained for the same compound a semiconducting band structure with a miniscule gap for the majority spin. Chantis \textit{et al.}~\cite{Chantis2007} applied a quasi-particle self-consistent G$W$ (QSG$W$) approach to various rare-earth monopnictides including GdN in the ferromagnetic state. They obtained a direct gap of \unit[0.46]{eV} for majority and \unit[1.48]{eV} for minority spin, which upon spin averaging agrees reasonably with the optical gap of \unit[1.31]{eV}  measured in experiment. However,  strong electronic correlations on localized 4$f$ shells  are a difficult case for weak-coupling perturbative approaches such as G$W$. 
Indeed, the QSG$W$ calculations overestimate the splitting between the occupied and unoccupied rare-earth 4$f$ states with an upper Hubbard band lying significantly higher in energy compared to DFT+$U$ calculations. The same problem regarding the treatment of 4$f$ states was observed in Ref.~\cite{Lambrecht2015}, where the  QSG$W$ approach has been applied to DyN, GdN and HoN.  

Strong local correlations on 4$f$ shells can be adequately described  within the non-perturbative  
DMFT approach~\cite{Metzner1989,Georges1992}. The combination of DFT with DMFT~\cite{Anisimov1997_1}, abbreviated as DFT+DMFT,  has been applied to numerous correlated materials \cite{Kotliar2006,Held2007} including rare-earth  metals~\cite{Locht2016},  monopnictides~\cite{Pourovskii2009,Peters2014} and monochalcogenides~\cite{Lebeque2005}. These works on lanthanide compounds employed  the quasi-atomic Hubbard-I approximation~\cite{hubbard_1}  neglecting hybridization of the 4$f$ states in the DMFT impurity problem.  This DFT+Hubbard-I approach is well suited to address compounds with quasi-atomic rare-earth 4$f$ shells, where it is able to capture multiplet effects absent in effective one-electron methods such as DFT+$U$. Moreover, in contrast to effective one-electron methods, DFT+Hubbard-I is  able to properly describe 4$f$ localization in the paramagnetic state. It is known from experiment that most $Ln$N exhibit ferromagnetic order at low temperatures ~\cite{Trodahl2007_GdN,Meyer2008,Lambrecht2013,Hewett2019_NdN,Hewett2020_DyN}.
However, their Curie temperatures are quite low---with a maximum $T_C=\unit[68]{K}$ for GdN~\cite{Granville2006}---, so that at room temperature they are actually all paramagnets.

Merits and drawbacks of various approaches (standard DFT, DFT+$U$, DFT+Hubbard-I) for TbN have been evaluated by Ref.~\cite{Peters2014}. While only DFT+Hubbard-I is able to capture the expected atomic multiplets, none of the employed approaches predicts a semiconducting gap, as apparent from the computed metallic spectral functions~\cite{Peters2014}. This is due to the fact that DFT+Hubbard-I includes only local electronic correlations,
so it does not improve on the underestimation of semiconducting band gaps.
Thus, previous theoretical works on $Ln$N either corrected the semiconducting $pd$ gap by employing advanced approaches to non-local exchange or included local correlations on the 4$f$ shells within a DMFT framework. 
A comprehensive first principles study of $Ln$N taking into account both effects is still lacking.

In the present paper we tackle this problem by calculating the electronic structure  of the $Ln$N series using an {\it ab initio} method that includes nonlocal exchange  through the modified Becke-Johnson (mBJ~\cite{tran_mbj_original,koller_mbj_2011}) exchange potential while local 4$f$ correlations are simultaneously included with DMFT in the Hubbard-I approximation. The on-site Coulomb repulsion $U$ for 4$f$ shells is calculated by  a constrained DFT+Hubbard-I technique, which we describe in detail in the methods section.   This methodology termed mBJ+Hubbard-I has very recently been applied to the rare-earth fluorosulfides $Ln$SF~\cite{Galler2021_LnSF} and rare-earth sesquioxides $Ln_2$O$_3$~\cite{Boust2021}. Here, we focus on the paramagnetic phase of  $Ln$N, which has been scarcely explored in previous theoretical works, and analyze in detail the evolution of the $Ln$N electronic structure along the series.

The structure of the article is the following: first, we outline our computational mBJ+Hubbard-I framework in Sec.~\ref{methods}, with a particular focus on the \textit{ab initio} computation of the screened Coulomb interactions $U$ within a  constrained DFT+Hubbard-I approach.  Our results, including an overview of computed optical gaps and \textbf{k}-integrated spectral functions for all investigated $Ln$N from PrN to ErN, are described in Sec.~\ref{results}. In particular, our analysis is focused on several compounds---NdN, SmN, TbN and HoN---that have been subject to intensive discussions in the recent literature.
Our conclusions are presented in Sec.~\ref{conclusion}.

\section{Methods}
\label{methods}

\subsection{The mBJ+Hubbard-I approach}
We start from a charge-self-consistent DFT+DMFT calculation~\cite{wien2k,Aichhorn2009,Aichhorn2011,triqs_cpc_2015,triqs_dfttools} of the target rare-earth mononitrides $Ln$N  ($Ln$ = Pr, Nd, Sm, Gd, Tb, Dy, Ho, Er).  All these compounds have localized $Ln$ 4$f$ shells, hence the quasi-atomic Hubbard-I  approximation can be employed as DMFT impurity solver.  In rare-earth semiconductors, corrections to the Hubbard-I solution due to hybridization effects are  most significant  for occupied 4$f$ states located  inside the semiconducting $pd$ gap~\cite{Boust2021}; this situation does not occur in the $Ln$N systems we consider. Since hybridization corrections are expected to be small otherwise \cite{Galler2021_LnSF,Boust2021}, we neglect them in the present work.

We construct projective Wannier functions~\cite{Aichhorn2009} to represent the subspace of the correlated $Ln$ 4$f$ states using the Kohn-Sham eigenstates enclosed by an energy window $[-9.5:13.6]$~eV around the Fermi level. We employ a fully rotationally-invariant screened Coulomb interaction in our calculations. 

After convergence of the self-consistent DFT+Hubbard-I calculations, we run an additional DFT cycle employing the Tran-Blaha modified Becke-Johnson (mBJ) potential~\cite{tran_mbj_original,koller_mbj_2011}, as  implemented in the wien2k~\cite{wien2k} program package. Such a perturbative use of the mBJ potential is appropriate since, strictly speaking, the mBJ potential is not variational, meaning, not derived from the minimization of a total-energy functional. It has further been shown that self-consistent mBJ calculations often exhibit convergence problems, while a perturbative use thereof can yield more reliable values for semiconducting band gaps~\cite{hong_mbj_2013}. For more details regarding our mBJ+Hubbard-I approach we refer the reader to Ref.~\cite{Boust2021}.

In the rare-earth 4$f$ quantum impurity problem we include all fourteen 4$f$ orbitals, which form two manifolds---$j = 5/2$ and $j = 7/2$---split by the spin-orbit coupling; additional smaller splittings within each manifold arise due to the crystal field. We employ the fully-localized-limit double-counting correction in the atomic limit~\cite{Pourovskii2007}, i.e. $\Sigma_{DC} = U (N-0.5)-J (0.5N - 0.5)$ with the corresponding $Ln^{3+}$ nominal atomic occupancies $N$. 
All calculations are carried out for a temperature of \unit[290]{K}. 

\subsection{Ab initio calculation of the screened Coulomb interaction U}

The Coulomb repulsion on an $f$ shell is determined by four Slater parameters $F^0$, $F^2$, $F^4$ and $F^6$.  In strongly localized $Ln$ 4$f$ shells, the parameters $F^2$, $F^4$ and $F^6$ are well known to exhibit virtually no material dependence.  Within the spherical approximation, which is reliable for 4$f$ shells, $F^2$, $F^4$ and $F^6$ are given by a single parameter---the Hund's rule coupling $J_H$. We thus employ for $J_H$  the values extracted from optical measurements of Ref.~\cite{carnall_lanthanides_1989}.
The parameter $F^0 \equiv U$, in contrast, is strongly reduced from its atomic value by screening processes in solids; this screening is determined by the electronic structure of a given compound.

 In order to calculate $U$ for the $Ln$N series, we developed a constrained DFT+Hubbard-I (cDFT+Hubbard-I) technique  based on the well-known constrained-DFT (cDFT) method for calculating the screened Coulomb interaction. In standard cDFT \cite{Dederichs1984,Hybertsen1989,Gunnarsson1989,Anisimov1991,Cococcioni2005}, the target shell occupancy (e.g. 4$f$) on a chosen  site is constrained to a predefined value with the rest of the electrons allowed to screen it. By increasing or decreasing this constrained occupancy one can evaluate the cost in interaction energy of placing electrons on the target shell, i.e. the parameter $U$. Its value is extracted either from the change in total energy due to the variation in occupancy or from the corresponding energy shift of the target band.  The cDFT method is not free of uncertainties, related to separating out the kinetic and interaction energy contributions to the total energy cost~\cite{Anisimov1991,Cococcioni2005} as well as to the  inter-site  interaction between constrained shells. To reduce the impact of the latter, cDFT calculations are typically carried out for reasonably large supercells so that constrained sites are well separated  in the real space. 

For rare-earth semiconductors, the applicability of standard cDFT is questionable. The only metallic bands in their DFT electronic structure are of $Ln$ 4$f$ character (in contrast to the case of $Ln$ metals, where 6$s$ and 5$d$ metallic bands are also present and thus cDFT performs reasonably well~\cite{Anisimov1991}). In cDFT, the constrained charge on a chosen $Ln$ site will thus  be  screened by metallic 4$f$ states. However, the metallic 4$f$  bands with their contribution to screening are an artifact of DFT, since in reality the 4$f$ electrons on $Ln$ ions are essentially quasiatomic. A related problem is that the $pd$ semiconducting bands in DFT are strongly impacted by hybridization with the 4$f$ metallic band in the middle of the $pd$ gap, with the gap magnitude enhanced to about \unit[3]{eV} instead of being on the verge of semi-metallicity (see Appendix). The DFT band structure for the $Ln$N series is thus very far from being realistic.

In our cDFT+Hubbard-I approach, we instead exploit the natural ability of the Hubbard-I approximation to constrain the 4$f$ shell occupancy to chosen integer values while keeping the shell localized. The DMFT impurity problem within this approximation is reduced to diagonalization of the 4$f$ Hamiltonian~\cite{Lichtenstein_LDApp}:
\begin{equation}\label{eq:H_at_full}
\hat{H}_{at}=\hat{H}_{1el}+\hat{H}_U=\sum_{\Lambda\Lambda'}\epsilon_{\Lambda\Lambda'}f^{\dagger}_{\Lambda} f_{\Lambda'}+\hat{H}_U, 
\end{equation}
where $f_{\Lambda'}$ ($f^{\dagger}_{\Lambda}$) is the creation (annihilation) operator for the $Ln$ 4$f$ orbital labeled by the combined spin-orbital index $\Lambda\equiv m\sigma$ (with $m$ and $\sigma$ being the magnetic and spin quantum numbers, respectively), $\hat{H}_U$ is the on-site Coulomb repulsion. The one-electron level-position matrix $\hat{\epsilon}$ reads:
\begin{equation}\label{H1el_HI}
\hat{\epsilon}=-\mu+\hat{H}_{KS}^{ff} -V
\end{equation}
where $\mu$ is the chemical potential, $\hat{H}_{KS}^{ff}=\sum_{\vk \in BZ}\hat{P}_{\vk} H_{KS}^{\vk}\hat{P}_{\vk}^{\dagger}$ is the Kohn-Sham Hamiltonian projected to the basis of 4$f$ Wannier orbitals 
and summed over the Brillouin zone, $\hat{P}_{\vk}$ is the corresponding projector between the Kohn-Sham and Wannier spaces \cite{Aichhorn2009}. The shift $V$ in the standard DFT+Hubbard-I should be equal to the double counting correction term  $\Sigma_{\mathrm{DC}}$ as can be shown by a high-frequency expansion of the DMFT bath Green's function~\cite{Pourovskii2007}.   In the cDFT+Hubbard-I approach we instead treat $V$ as a uniform potential applied to the 4$f$ shell on a given site and choose its value to constrain the occupancy.  The atomic ground-state 4$f$ occupancy $N^{at}_{GS}$  as a function of $V$ forms an upward staircase with each plateau corresponding to an integer value $N^{at}_{GS}$ between 0 and 14. The occupancy  $N_{GS}$ of a correlated shell in DFT+Hubbard-I is calculated from  the local Green's function evaluated with the corresponding atomic self-energy inserted at all correlated sites. For  strongly localized states  $N_{GS}$ vs. $V$ exhibits qualitatively the same behavior as $N^{at}_{GS}$, apart from the step-like transitions smoothened  by hybridization effects.  Hence, the value of $V$ to obtain a required integer occupancy of $N_{GS}$ can be easily found.

The cDFT+Hubbard-I calculation is carried out for a $Ln$N supercell with the on-site potential $V$ on two chosen $Ln$ sites tuned to constrain their 4$f$ occupancy to $N_{GS}+1$ and $N_{GS}-1$, correspondingly, where $N_{GS}$ is the ground-state 4$f$  occupancy for a given $Ln$ ion. For other $Ln$ ions in the supercell we apply the standard Hubbard-I with $V=\Sigma_{DC}$,  their 4$f$ occupancy remains equal to $N_{GS}$. The value of $U_{cHI}$ defining the on-site 4$f$ Coulomb repulsion $\hat{H}_U$ in these cDFT+Hubbard-I  calculations can be chosen quite arbitrarily, though it needs to be sufficiently large to have well defined plateaus for integer    4$f$ occupancies as a function of $V$. We fixed it at \unit[10]{eV}.

Once the cDFT+Hubbard-I calculations converge, the value of $U$ can be extracted from the difference of averaged $ff$ blocks of the Kohn-Sham Hamiltonian  $\langle \hat{H}_{KS}^{ff} \rangle$ between the two sites with constrained occupancies. Namely, the orbital/spin average of $\langle \hat{H}_{KS} ^{ff}\rangle=\frac{1}{14}\sum_{\Lambda} \left[\hat{H}_{KS} ^{ff}\right]_{\Lambda\Lambda}$ for a given occupancy $N$ reads
\begin{equation}
\langle \hat{H}_{KS}^{ff}\rangle_N=\langle \hat{H}^{0ff}_{KS}\rangle+U(N-1/2)-J_H(N/2-1/2),
\end{equation}
where $\langle \hat{H}^{0ff}_{KS}\rangle$ is this average excluding the contribution of the intra-shell Coulomb repulsion, for the latter we assume the  fully-localized limit form as given by the second and third terms on the right hand side. Since $\langle \hat{H}^{0ff}_{KS}\rangle$  does not depend on the 4$f$ shell occupancy, one finds:
\begin{equation}
U=\frac{1}{2}\left[\langle \hat{H}_{KS}^{ff}\rangle_{N+1}-\langle \hat{H}_{KS}^{ff}\rangle_{N-1}+J_H\right],
\end{equation}
which is the equation we used to extract $U$ from our cDFT+Hubbard-I results.

The present technique thus evaluates $U$  for a realistic electronic structure of  the $Ln$ semiconductors, in which the 4$f$ states are localized by the Hubbard interaction and do not contribute to any metallic screening of the constrained charge. 

We also note that the average $ff$ block of the Kohn-Sham Hamiltonian $\langle \hat{H}_{KS}^{ff}\rangle_N$ is not equal to the  centreweight of the corresponding 4$f$ band. The latter is impacted by hybridization effects, leading to the complex problem of removing these (kinetic energy) effects from the cDFT estimation for $U$ \cite{Cococcioni2005}. In contrast, $\langle \hat{H}_{KS}^{ff} \rangle_N$ gives the  4$f$ band position once hybridization of 4$f$ with other states, within the energy window, is suppressed. Since in our calculations we employ a large window including  all relevant valence bands, our estimation for $U$ is essentially free from any admixture of hybridization effects.

Our cDFT calculations were carried out for a 32 atoms supercell with 15 $\vk$-points in the irreducible Brillouin zone. Consistently with the mBJ+Hubbard-I calculations for $Ln$N, the energy window for the Wannier projection was chosen to be $[-9.5:13.6]$~eV around the Fermi level.

\section{Results and discussion}
\label{results}

\begin{figure*}[htbp]
    \begin{center}
	\includegraphics[width=0.65\columnwidth]{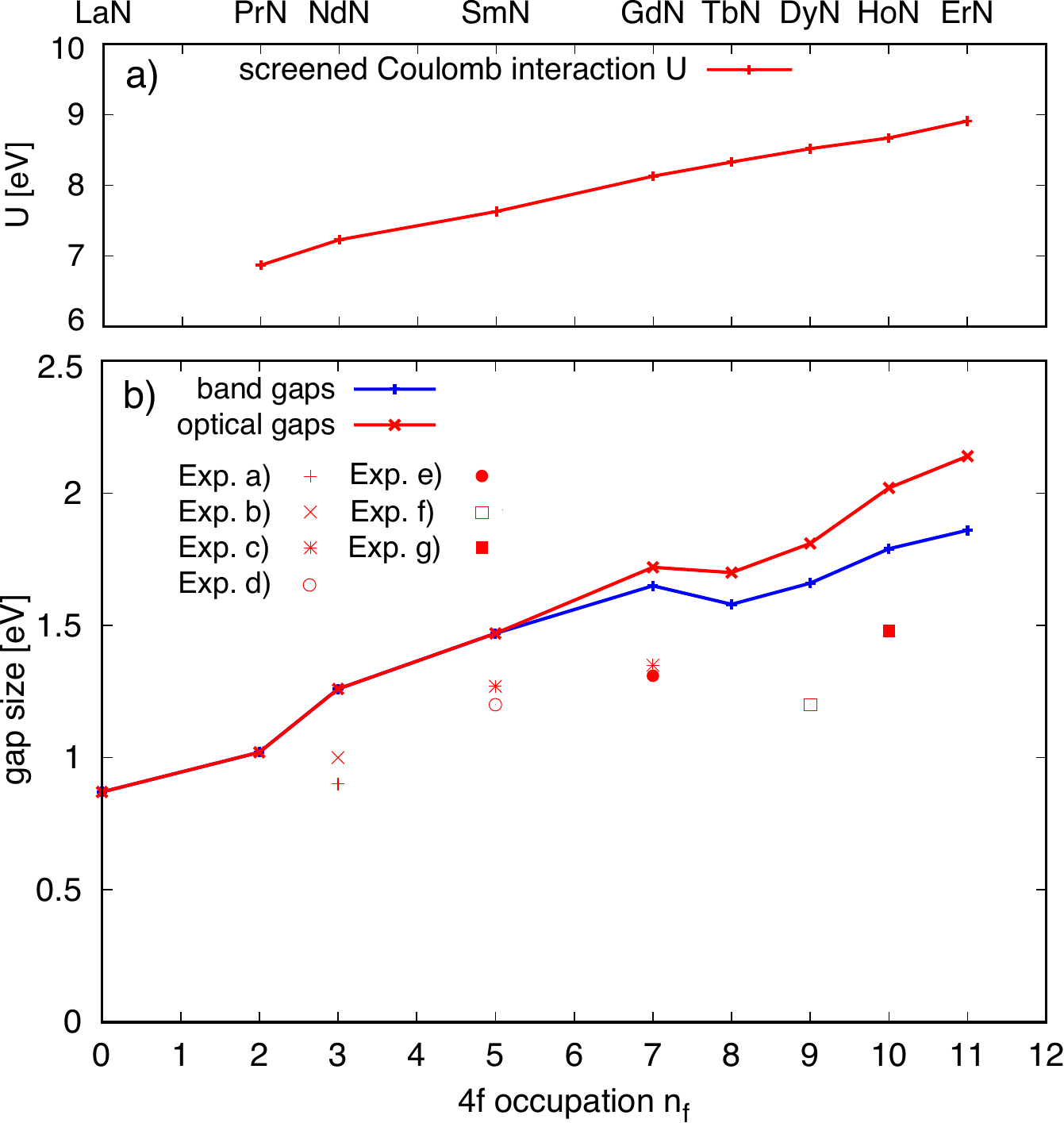}
	\caption{a) Screened on-site Coulomb interaction $U$ along the $Ln$N series, calculated from cDFT+Hubbard-I. b) Optical gap (red lines-points) and band gap (blue lines-points), calculated from mBJ+Hubbard-I. For heavier rare-earths starting with $Ln=$ Gd, the band gap becomes indirect and is slightly smaller than the direct optical gap at $X$. For comparison, experimental values for the optical gap are shown (red symbols), adapted from Refs. a)~\cite{Trodahl2016}, b)~\cite{Hewett2019_NdN}, c)~\cite{Hewett2019_SmN}, d)~\cite{Azeem2018_SmN}, e)~\cite{Trodahl2007_GdN}, f)~\cite{Hewett2020_DyN} and g)~\cite{Brown2012}. \label{fig:U_and_gaps}} 
	 \end{center}
\end{figure*}

\begin{figure*}[htbp]
    \begin{center}
	\includegraphics[width=1.0\columnwidth]{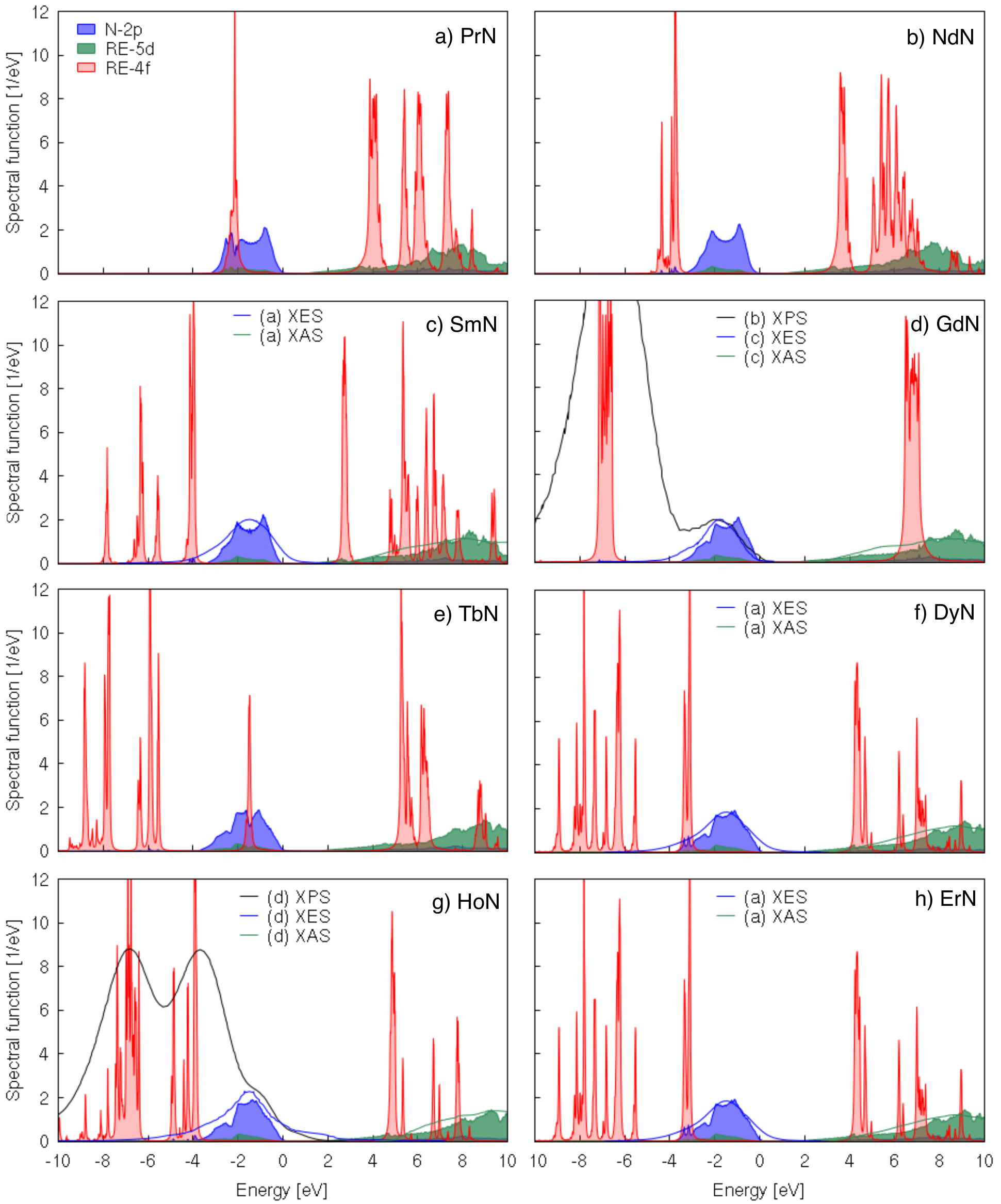}
	\end{center}
	\caption{Spectral functions of $Ln$N ($Ln$ 4$f$ in red, N 2$p$ in blue, $Ln$ 5$d$ in green).
	For comparison, available experimental  XPS (solid black lines), XES (blue lines)  and XAS (green lines) spectra are shown. Experimental spectra in arbitrary units reproduced from a) Ref.~\cite{Preston2007}, b) Ref.~\cite{Leuenberger2005}, c) Ref.~\cite{Preston2010_GdN} and d) Ref.~\cite{Brown2012}.}
	\label{fig:dos}
	
\end{figure*}

All rare-earth mononitrides $Ln$N crystallize in the simple fcc rocksalt structure with decreasing lattice constant along the series. In our calculations, we have employed experimental lattice constants as summarized in Ref.~\cite{Lambrecht2013}. We have performed calculations for eight members of the $Ln$N series with $Ln$= Pr, Nd, Sm, Gd, Tb, Dy, Ho, Er. 

\subsection{Coulomb interaction $U$ along the $Ln$N series}

We first present the evolution of the Coulomb interaction parameter $U$ along the $Ln$ mononitride series, as obtained by the cDFT+Hubbard-I method. The value of $U$   increases along the series from \unit[6.87]{eV} for PrN to \unit[8.91]{eV} for ErN (Fig.~\ref{fig:U_and_gaps}a and Table~\ref{tab:gaps}). This effect is caused by  the corresponding enhancement in the $4$f shell localization manifested in the well-know reduction of the $Ln^{3+}$ ionic radius along the lanthanide series.  $U$ exhibits a quasi-linear trend as a function of the 4$f$ shell occupancy $n_f$,  with a slightly more rapid increase in the beginning of the series, from Pr to Nd.

Previously, the $U$ values along this series have been estimated   by Larson {\it et al.}~\cite{Larson2007}, who first computed the bare unscreened value of $F^0$ for $Ln^{3+}$ ions and then applied a constant factor  to simulate its reduction by screening in $Ln$N compounds. The screening factor was extracted in Ref.~\cite{Larson2007} by comparing the bare Gd $F^0$  with the value of $U$ needed to align the  4$f$ band position in GdN calculated by LSDA+$U$ with that measured in photoemission. Their resulting $U$ values are about 10\% larger than ours (e.g., for GdN $U=\unit[9.2]{eV}$ as compared to our value of \unit[8.13]{eV}) and exhibit a somewhat more pronounced and non-monotonic increase along the series. 

The Hund's coupling $J_H$, listed in the second row of Table~\ref{tab:gaps}, is also progressively growing along the series, with values reaching from $J_H=\unit[0.73]{eV}$ in PrN to $J_H=\unit[1.05]{eV}$ in ErN. As noted above, $J_H$ for the 4$f$ shell is barely affected by the crystalline environment, the reported values for $J_H$ were extracted from optical measurements of $F^2$, $F^4$ and $F^6$ in rare-earth metals~\cite{carnall_lanthanides_1989}.

\subsection{Trends in the $Ln$N electronic structure}

The calculated $Ln$N \textbf{k}-integrated spectral functions 
are shown in Fig.~\ref{fig:dos}  together with available experimental x-ray photoemission (XPS), x-ray emission (XES) and x-ray absorption (XAS) spectra.
As expected, the topmost valence bands are mainly of N 2$p$ character and the conduction bands of $Ln$ 5$d$ character. Our calculations predict a semiconducting electronic structure for all $Ln$N, with valence N 2$p$  and conduction $Ln$ 5$d$ bands never overlapping in energy. 
The $Ln$ 4$f$ states are split into an occupied lower Hubbard band (LHB) and unoccupied upper Hubbard band (UHB).  
The occupied 4$f$ states progressively shift to lower energies along the series, from PrN to ErN; this evolution is due to the well-known increase of the 4$f$ binding energy along the rare-earth series. 
The Hubbard bands further split into multiple peaks due to transitions between quasiatomic multiplets. This multiplet splitting is a characteristic feature of $Ln$ 4$f$ states weakly hybridizing with the N 2$p$ and $Ln$ 5$d$ states and  thus keeping a quasi-atomic character in solids.  The multiplet splitting is characteristic for each $Ln$ element and is absent in the half-filled Gd 4$f$ shell which displays two sharp Hubbard bands at \unit[-7]{} and \unit[+7]{eV}, respectively. The position of the GdN LHB  agrees well with the experimental XPS spectrum of Ref.~\cite{Leuenberger2005}. The 4$f$ UHB is about 14~eV above the LHB, in agreement with the magnitude of effective $U_{eff}=U+6J_H$ for a half-filled $f$ shell.

Another, smaller peak around  \unit[-2]{eV} is clearly visible in the GdN XPS spectrum and
is identified by our calculations as the N 2$p$ band. 
The N 2$p$ states are further resolved in N K-edge XES data, while the unoccupied $Ln$ 5$d$ states are probed by XAS. They both well match our calculated spectrum, thus validating our methodology in the case of Gd mononitride. GdN  is by far the most investigated compound of the $Ln$N series, both experimentally and especially theoretically, since its electronic structure can---due to its half-filled 4$f$ shell and the absence of multiplet effects-- be qualitatively captured without employing DMFT. Hence, we will in the following rather concentrate on other, less explored compounds.

Unfortunately, there are few  experimental spectra available for the rest of the Ln$N$ series. For the existing ones, i.e. an XPS spectrum for HoN~\cite{Brown2012} as well as XES and XAS spectra for SmN~\cite{Preston2007}, DyN~\cite{Preston2007} and HoN~\cite{Brown2012}, our {\it ab initio} electronic spectra agree very well with experiment, as can be  seen in Fig.~\ref{fig:dos}. In particular, a valence-band XPS is available only for HoN; as one sees, the position of the occupied 4$f$ band and its splitting into two well separated manifolds is very well reproduced by our calculations. 

\begin{table}[tbp]
\centering
\begin{tabular}{c c c c c c c c c} 
 \hline\hline
  & PrN & NdN & SmN & GdN & TbN & DyN & HoN & ErN \\ \hline
  $U$ & 6.87 & 7.23 & 7.63 & 8.13 & 8.33 & 8.52 & 8.67 & 8.91 \\ 
  $J_H$ & 0.73 & 0.77 & 0.85 & 0.92 & 0.95 & 0.98 & 1.01 & 1.05 \\ 
  \hline
   \multicolumn{9}{c}{Optical gap} \\
  \small{This work} & 1.02 & 1.26 & 1.47 & 1.72 & 1.70 & 1.81 & 2.02 & 2.14 \\ 
  \multirow{2}*{\small{Exp}}& &0.9\textsuperscript{\cite{Trodahl2016}}  & 1.2\textsuperscript{\cite{Azeem2018_SmN}}  & 1.31\textsuperscript{\cite{Trodahl2007_GdN}}  &   & \multirow{2}*{1.2\textsuperscript{\cite{Hewett2020_DyN,Azeem2013_DyN}}}  & \multirow{2}*{1.48\textsuperscript{\cite{Brown2012}}}  &  \\
  & & 1.0\textsuperscript{\cite{Hewett2019_NdN}} &  1.27\textsuperscript{\cite{Hewett2019_SmN}} &  1.35\textsuperscript{\cite{Hewett2019_SmN}} & & & & \\ \hline\hline  
 \end{tabular}
 \caption{Computed values of the screened Coulomb interaction $U$ and optical gaps in the rare-earth mononitrides. Tabulated are further the experimental values of the Hund's coupling $J$, employed in our calculations and extracted from Ref.~\cite{carnall_lanthanides_1989} (for rare-earth metals). For comparison, experimental data of optical gaps are listed, taken from several studies performed during the last 20 years.} 
\label{tab:gaps} 
\end{table}

In Fig.~\ref{fig:U_and_gaps}b we display our calculated values for the optical gap along the $Ln$N series, see also Table~\ref{tab:gaps}. These values are extracted from the $Ln$N $\textbf{k}$-resolved spectral functions (some of them are shown in Figs.~\ref{fig:bands_nd_sm} and \ref{fig:bands_tb_ho}).  Despite their progressive downward shift and changing shape along the series, the $Ln$ 4$f$ states never touch the bottom of  the conduction band or the top of valence band. Therefore, for all investigated compounds the direct optical gap at $X$ is formed between the N 2$p$ and the $Ln$ 5$d$ states. 
The optical gap progressively increases from \unit[1.02]{eV} in PrN to \unit[2.14]{eV} in ErN. A comparison to available experimental data (Fig.~\ref{fig:U_and_gaps}b and Table~\ref{tab:gaps}) reveals a qualitatively similar, though less pronounced trend  in experiment. Our calculations predict an overall steady increase of the optical gap along the series, with two noticeable peculiarities, i.e. a kink from Pr to Nd and  a plateau between Gd and Dy. A similar non-trivial gap evolution vs. $n_f$  is observed in experiment, with the value of the optical gap even slightly decreasing from Gd to Dy.  The gaps' absolute value is slightly overestimated by our mBJ+Hubbard-I methodology. For example, in the experimentally well investigated compound GdN, the difference between theoretically predicted and experimentally measured gap size is around 0.3 eV. Such a systematic overestimation seems to be a general feature of the mBJ potential applied to $d$-electron conduction states, as previously observed in the case of $d^0$ titanates~\cite{hong_mbj_2013}. 

Interestingly, our calculations predict the band gap to become indirect in the second half of the series due to the maximum of the valence band shifting from the $X$ to $\Gamma$ point (cf. Figs.~\ref{fig:bands_nd_sm} and \ref{fig:bands_tb_ho}). Such a shift is absent in the pure DFT band structure, though the N 2$p$ band maximum at $X$ becomes more shallow for heavy $Ln$, see Appendix Fig.\ref{fig:KS_bands}. With the localized 4$f$ nature  properly included in DFT+Hubbard-I, the $pd$ gap drastically shrinks once the metallic 4$f$ bands are removed by the on-site Coulomb interaction. This gap reduction results in a stronger downward shift of the 2$p$ band maximum at $X$ by $pd$ hybridization. The hybridization shift at $X$ grows along the series due to the reducing cell volume with the corresponding increase in $pd$ hopping.  This results in the band gap becoming indirect $\Gamma-X$ starting from GdN.

This trend was previously only roughly captured by LSDA+$U$ calculations of Larson {\it et al.}~\cite{Larson2007}. In the case of completely empty (LaN) and filled (LuN) 4$f$ bands they obtained direct and indirect band gaps, respectively. However, for all considered compounds with a fractional 4$f$ occupancy, from PrN to YbN, an indirect gap was predicted for the paramagnetic phase. Therefore, a sharp transition to an indirect gap was predicted once the 4$f$ band becomes partially occupied; the calculated band gap exhibits no clear trend along the series from Pr to Yb.  Our  calculations directly treating the paramagnetic phase rather predict a smooth evolution of the $pd$ gap along the series.

We will now supplement the general picture presented above with a more detailed discussion of four selected compounds -- NdN, SmN, TbN, and HoN -- that exemplify the electronic structure  evolution along the $Ln$N series.

\subsection{NdN}

\begin{figure*}[htbp]
	\includegraphics[width=1.0\columnwidth]{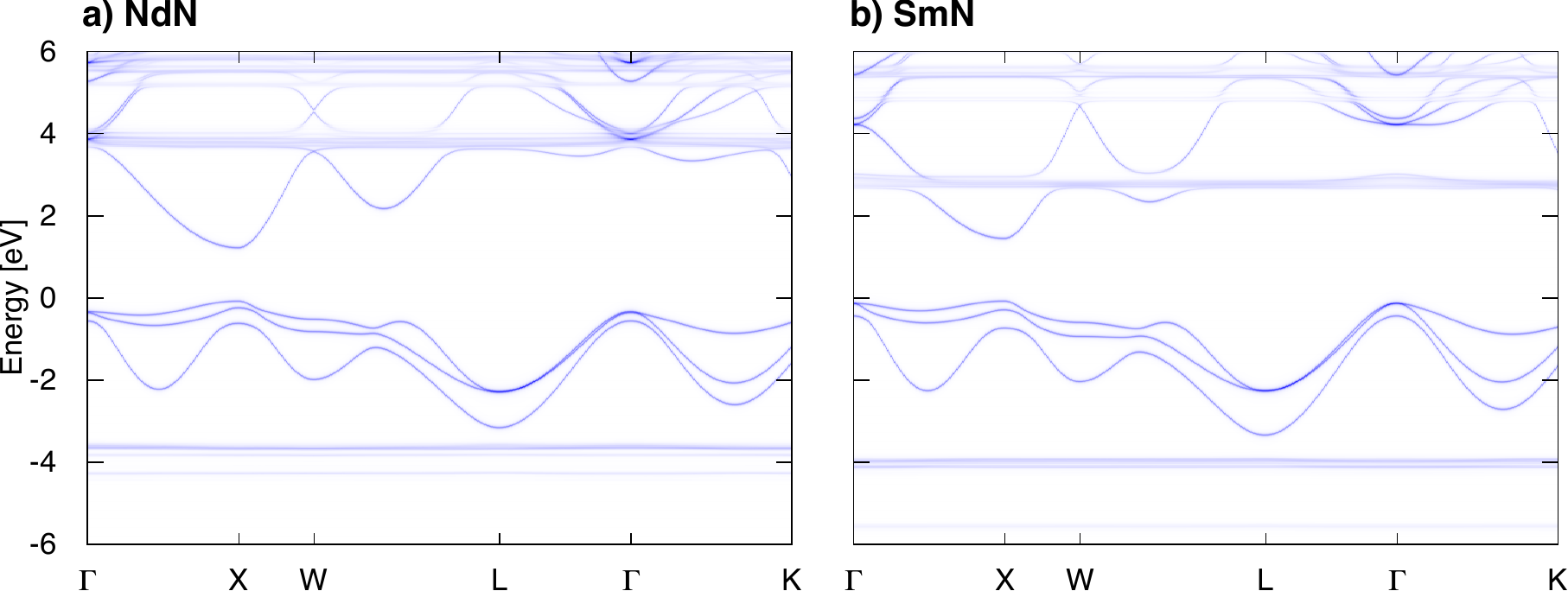}
	\caption{\textbf{k}-resolved spectral functions of a) NdN and b) SmN, computed within our mBJ+Hubbard-I approach.}
	\label{fig:bands_nd_sm}
\end{figure*}

In Fig.~\ref{fig:bands_nd_sm}a we show the computed \textbf{k}-resolved spectral function of NdN in the vicinity of the Fermi energy, i.e. from \unit[-6]{} to  \unit[6]{eV}, while the overall \textbf{k}-summed spectral function in a larger energy range has been presented in Fig.~\ref{fig:dos}b.  
Like in all investigated $Ln$N,
we  obtain a $Ln^{3+}$ configuration, resulting in three electrons in the Nd 4$f$ shell. As clearly visible from the \textbf{k}-integrated spectral function in Fig.~\ref{fig:dos}b, the Nd 4$f$ states form a rather sharp lower Hubbard band centered around \unit[-4]{eV}, while the unoccupied upper Hubbard band spreads from \unit[3]{} to almost \unit[10]{eV} due to the effect of multiplet splittings.  
The band  gap of \unit[1.26]{eV} is direct and located at the $X$ point, as one sees from the NdN \textbf{k}-resolved spectral function (Fig.~\ref{fig:bands_nd_sm}a).
The calculated gap magnitude, \unit[1.26]{eV}, is slightly overestimated by our calculations, since optical measurements in Refs.~\cite{Trodahl2016,Hewett2019_NdN} reported \unit[0.9-1.0]{eV}.  This is consistent with the systematic overestimation of $pd$ gaps by the mBJ potential, as noted above. In spite of this overestimation, our theoretical value of the NdN band gap still agrees  with experiment significantly better than previous calculations.
 LSDA+$U$ calculations in Ref.~\cite{Larson2007} reported an (indirect, majority-spin) gap of only \unit[0.3]{eV} in NdN, while earlier calculations predicted a half-metallic state~\cite{Aerts2004}. 
According to our calculations, paramagnetic NdN is clearly a $pd$ semiconductor, in agreement with transport measurements~\cite{Trodahl2016}. 
The unoccupied 4$f$ states are located approximately \unit[2]{eV} above the conduction band minimum. They weakly hybridize with the conduction band, but do not form the conduction band minimum, as suspected recently in Ref.~\cite{Hewett2019_NdN}. We note, however, that our theoretical predictions are not at odds  with the measurement of the optical conductivity in Ref.~\cite{Hewett2019_NdN}. An earlier onset of the conductivity in NdN of around \unit[0.2]{eV} compared to GdN, can be explained not only by the presence of additional states, i.e. the rare-earth 4$f$ states, at the bottom of the conduction band, but also by an increase of the $pd$ gap along the $Ln$N series, as predicted by our calculations.

\subsection{SmN}
Resistivity measurements clearly indicate a semiconducting nature for SmN~\cite{Preston2007}, in the paramagnetic as well as in the magnetically ordered state below \unit[20-27]{K}~\cite{Preston2007,Meyer2008}.
However, previous studies within LSDA+$U$~\cite{Larson2007,Preston2007,Morari2015} predicted a zero gap and varying position of the Sm 4$f$ states depending on the choice of $U$.  
Our calculated \textbf{k}-integrated spectral function of SmN is presented in Fig.~\ref{fig:dos}c, while a zoom into the \textbf{k}-resolved spectrum around the Fermi energy is provided in Fig.~\ref{fig:bands_nd_sm}b.   Due to an increase in the 4$f$ binding energy, the occupied 4$f$ states in SmN lie lower in energy compared to NdN. They are located in the range from \unit[-8]{} to \unit[-4]{eV} and   split into several sharp peaks by multiplet effects (\ref{fig:dos}c). The unoccupied upper Hubbard band extends between \unit[2.5-10]{eV}, above the valence band maximum. For SmN, there are N K-edge x-ray emission (XES) and absorption (XAS) spectra available in Ref.~\cite{Preston2007}, which probe the occupied N 2$p$ and Sm 5$d$ states, respectively. They are shown in Fig.~\ref{fig:dos}c.   Overall, the shape of experimental and theoretical spectra agrees well, though the energy axis position of the XES and XAS spectra, which cannot reliably be probed in experiment, has been adjusted to match our theoretical spectrum. Unfortunately, an x-ray photoemssion  spectrum (XPS), which would probe the total occupied spectrum including the rare-earth 4$f$ states, is not yet available for SmN.

The \textbf{k}-integrated spectral function in Fig.~\ref{fig:dos}c could give the impression that the Sm 4$f$ states form the conduction band minimum. However, this is not the case, as can be seen from the \textbf{k}-resolved spectrum in Fig.~\ref{fig:bands_nd_sm}b. 
The lowest-lying unoccupied 4$f$ states hybridize with the Sm 5$d$ states and are closer to the conduction minimum at $X$ than in NdN, but still around \unit[1]{eV} higher in energy. The direct band gap in SmN is again formed between the occupied N 2$p$ and the unoccupied Sm 5$d$ bands. This picture is at odds with the conclusion of  Ref.~\cite{Hewett2019_SmN}, who attributed a broad absorption feature centered around \unit[0.5]{eV} in the optical conductivity of SmN to the presence of Sm 4$f$ states in the semiconducting gap. However, this additional broad absorption feature is more likely an effect of nitrogen vacancies, which are known to induce in-gap states. The latter hypothesis was followed in Ref.~\cite{Azeem2019_SmN} and has carefully been studied by the authors of Ref.~\cite{Hewett2019_SmN} later on for a very similar absorption feature in DyN~\cite{Hewett2020_DyN}.  
According to precise measurements of the optical gap in SmN~\cite{Azeem2018_SmN,Hewett2019_SmN}, our theoretically predicted gap of \unit[1.47]{eV} seems to be overestimated by around \unit[0.2]{eV}.

\subsection{TbN}
Let us now turn to a rare-earth mononitride with a more than half-filled 4$f$ shell. 
 Tb$^{3+}$ has 8 electrons in the 4$f$ shell, i.e. one electron more compared to the half-filled case. The occupied 4$f$ states in TbN are split into a manifold between \unit[-10]{} and \unit[-5]{eV}, exhibiting further intrinsic multiplet splittings, and a single peak slightly above \unit[-2]{eV} located within the N 2$p$ band and weakly hybridizing with it  (Fig.~\ref{fig:dos}e).  The peak above \unit[-2]{eV} is associated with the transition by electron removal from the half-filled Hund's rule $^8S_{7/2}$ state. The low-lying 4$f$ manifold is due to excited states of the half-filled shell, which have much higher energy than the Hund's rule state.

A density of states (DOS) for TbN was previously calculated by Aetrs~{\it et al.}~\cite{Aerts2004} by a  self-interaction-corrected (SIC) LSDA  approach. This study predicted  occupied 4$f$ states below \unit[-13]{eV} and  unoccupied ones right above the Fermi energy, which does not seem to be plausible.  An LSDA+$U$ band structure for TbN was calculated by Larson {\it et al.}.~\cite{Larson2007}. 
They  found two competing solutions, one dominated by the material's crystal fields while the second one induced by atomic Hund's rules.  In our many-body mBJ+Hubbard-I  approach we do not encounter this ambiguity since it includes the local Coulomb interaction on the 4$f$ shell, as well as  crystal-fields effects and spin-orbit coupling on equal footing. For TbN, a Hubbard-I calculation has already been performed in Ref.~\cite{Peters2014}. As expected, the main features of our \textbf{k}-integrated spectral function  in Fig.~\ref{fig:dos}e, in particular the characteristic multiplet splitting of the Tb$^{3+}$ shell, agree with the spectrum obtained  in Ref~\cite{Peters2014}. The strength of the splittings and exact position of 4$f$ peaks, however, slightly differ due to the difference in the employed $U$ and $J_H$ values. The most striking difference is the absence of a band gap in Ref.~\cite{Peters2014}. This is understandable from the fact that in Ref.~\cite{Peters2014} Hubbard-I was combined with LDA, which usually underestimates semiconducting band gaps. While, to date, there is  no experimental data available for TbN, it seems 
unlikely that TbN, contrary to its neighbours GdN and DyN, does exhibit a zero bandgap. 

\begin{figure*}[htbp]
	\includegraphics[width=1.0\columnwidth]{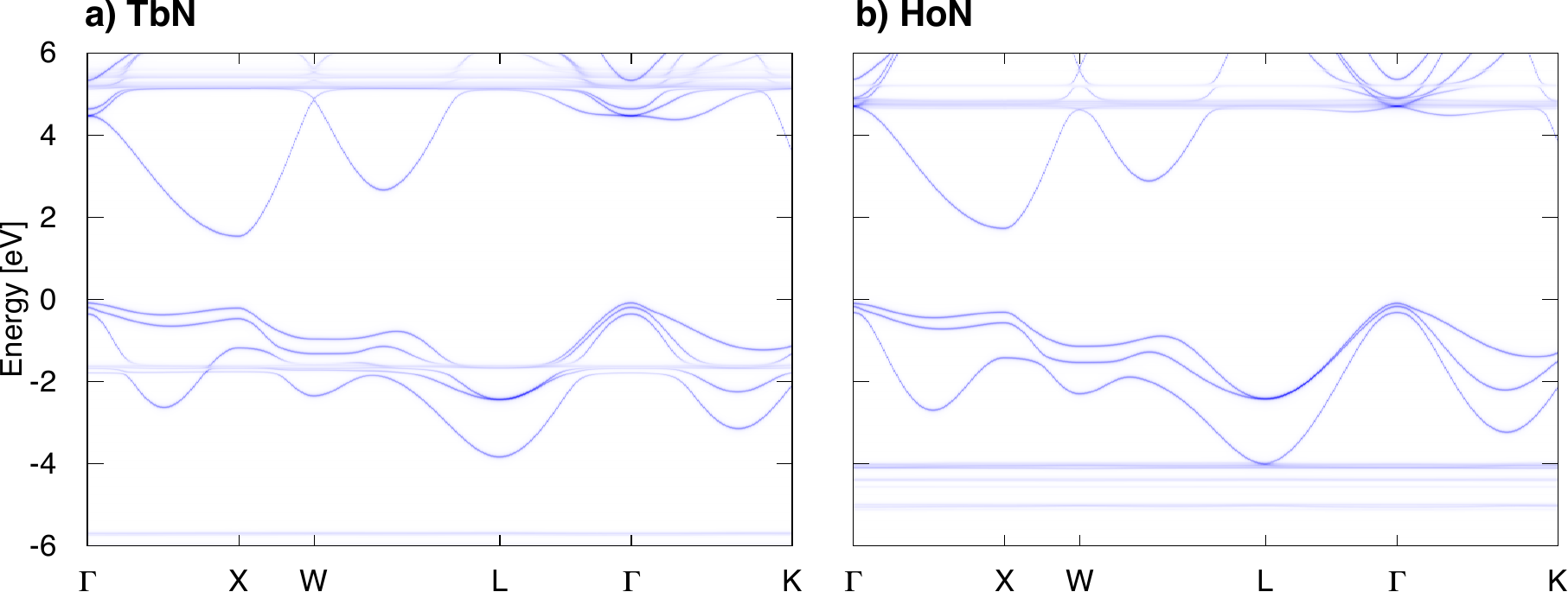}
	\caption{Computed \textbf{k}-resolved spectral functions of a) TbN and b) HoN.}
	\label{fig:bands_tb_ho}
\end{figure*}

Our mBJ-Hubbard-I approach predicts a direct optical gap of \unit[1.7]{eV} at $X$ for TbN, as can be seen from the \textbf{k}-resolved spectral function in Fig.~\ref{fig:bands_tb_ho}b. The indirect bandgap between the maximum of the valence band at $\Gamma$ and the minimum of the conduction band at $X$ is  \unit[0.15]{eV} smaller.

\subsection{HoN}
The Ho$^{3+}$ 4$f$ shell hosts 10 electrons. Like almost all $Ln$N, HoN exhibits ferromagnetic order below its Curie temperature of $T_C=\unit[18]{K}$~\cite{Yamamoto2004}, while above it is paramagnetic. The \textbf{k}-integrated spectral function of HoN is shown in Fig.~\ref{fig:dos}g.
 
Due to multiplet effects, the occupied 4$f$ states are split into two main manifolds centered around \unit[-7]{} and \unit[-4.5]{eV}, respectively. The upper Hubbard band, instead, spreads from \unit[4.5]{} to \unit[8]{eV}. The two main manifolds of the lower Hubbard band are clearly visible in the experimental x-ray photoemission (XPS) spectrum of Ref.~\cite{Brown2012}, which we have reproduced in Fig.~\ref{fig:dos}g. The  overlay with our theoretical spectrum shows an excellent agreement. The shoulder at the low binding energy side of the XPS spectrum matches the position and shape of the N 2$p$ states. The N 2$p$ states were further probed by N K-edge x-ray emission (XES) in Ref.~\cite{Brown2012}. The experimental XES spectrum shows rather broad N 2$p$ states with a low-energy tail extending to almost \unit[-8]{eV}, which led the authors of Ref.~\cite{Brown2012} to the conclusion of strong $p-f$ hybridization in HoN. In our calculations, the Ho 4$f$ states are well localized and the $p-f$ hybridization is not that pronounced. In the detailed experimental study of Ref.~\cite{Brown2012}, additionally, the unoccupied $Ln$ 5$d$ states were probed in x-ray absorption (XAS). The corresponding XAS spectrum is reproduced in Fig.~\ref{fig:dos}g.  

Our calculation predicts for HoN an optical gap of \unit[2.02]{eV} at $X$ (Fig.~\ref{fig:bands_tb_ho}b) . Again, the computed optical gap seems to be overestimated compared to its experimental value of \unit[1.48]{eV}~\cite{Brown2012}. Similar to TbN, the band gap in HoN is indirect with the valence band maximum located at $\Gamma$, and it is \unit[0.2]{eV} smaller than the optical gap.

\section{Conclusions}
\label{conclusion}
 We have computed the paramagnetic electronic structure of eight rare-earth mononitrides $Ln$N (with $Ln$ = Pr, Nd, Sm, Gd, Tb, Dy, Ho, Er) by employing an advanced first-principles mBJ+Hubbard-I approach. This approach
combines a quasi-atomic treatment of electronic correlations in the $Ln$ 4$f$ shell  with an improved description of semiconducting band gaps 
by the non-local mBJ exchange potential. The screened on-site Coulomb interaction in the 4$f$ shells is evaluated from first principles by a novel cDFT+Hubbard-I methodology that we describe in details in the present work.  
From the calculated spectral functions we obtained the evolution of the semiconducting band gap along the $Ln$N series as well as position and multiplet spitting of the $Ln$ 4$f$ states.  
A semiconducting gap of $pd$ type is predicted for all investigated compounds, with the gap magnitude ranging from \unit[1.02]{eV} in PrN to \unit[2.14]{eV} in ErN.  The band gap is direct in light lanthanide $Ln$N up to SmN; it becomes indirect $\Gamma-X$ for heavy $Ln$. The predicted evolution of the band gap magnitude along the series agrees qualitatively with available data from optical measurements; theoretical gap values are slightly  systematically overestimated. The calculated \textbf{k}-integrated spectral functions agree well with available spectra from x-ray photoemission (XPS), x-ray emission (XES) and x-ray absorption (XAS) spectroscopy. The present work 
provides a compendium of computed spectral functions of the rare-earth mononitrides $Ln$N, which can be useful for future experimental and theoretical investigations.  

Overall, the combination of the quasiatomic Hubbard-I treatment for local 4$f$ correlation with the non-local mBJ exchange potential has shown promising predictive capabilities 
for a variety of rare-earth based semiconductors including the rare-earth mononitrides $Ln$N, rare-earth fluorosulfides $Ln$SF~\cite{Galler2021_LnSF} and rare-earth sesquioxides $Ln_2$O$_3$~\cite{Boust2021}. 

\section*{Acknowledgements}
We thank Silke Biermann and James Boust for fruitful discussions. This work was supported by the Austrian Science Fund (FWF): Schr\"odinger fellowship J-4267. We thank the computer team at CPHT for support.

\section*{Appendix}
\begin{figure*}[htbp]
    \begin{center}
	\includegraphics[width=0.9\columnwidth]{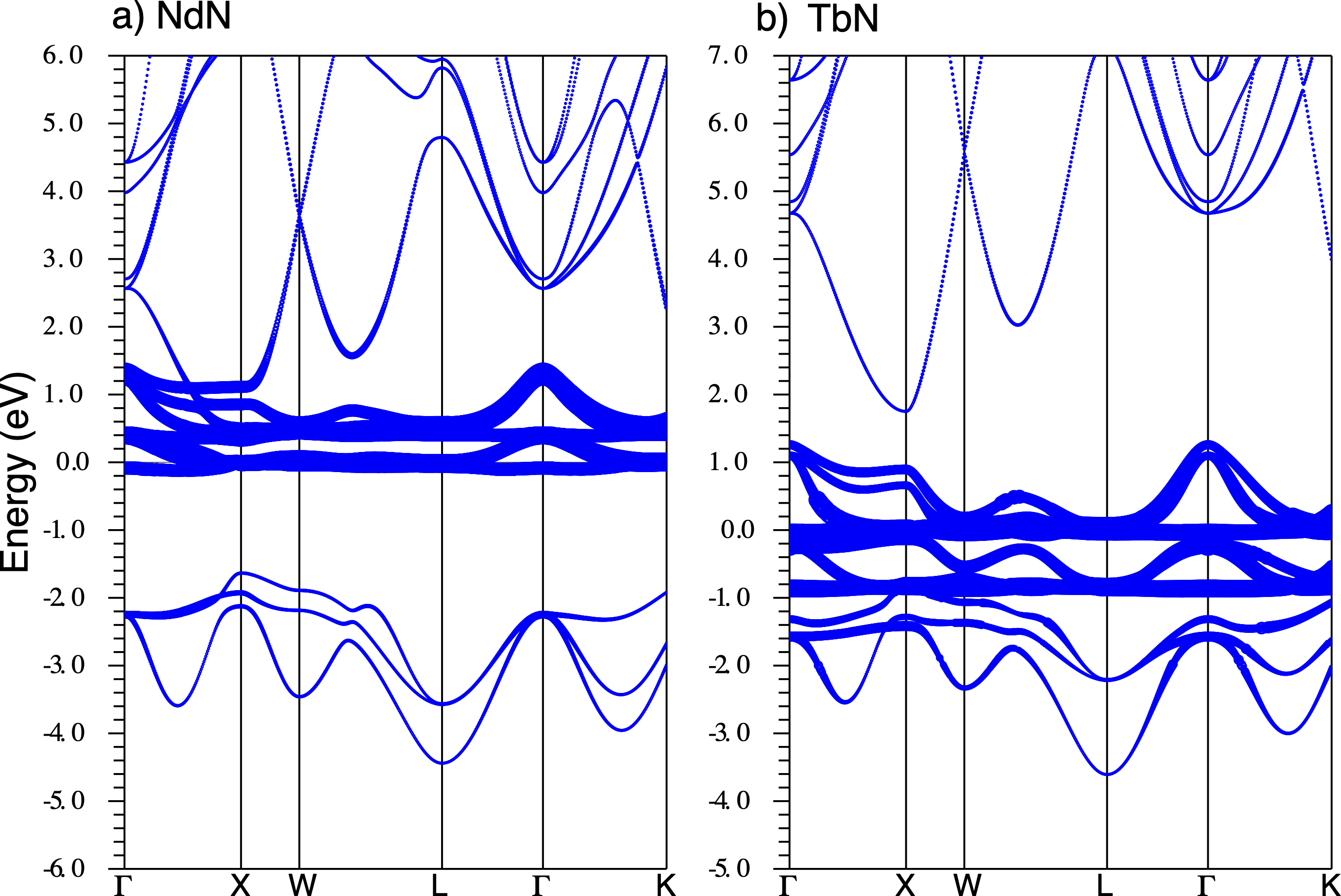}
	\caption{Kohn-Sham band structure of a) NdN and b) TbN. The fat-band plotting indicates the 4$f$ character of the bands. }
	\label{fig:KS_bands}
	\end{center}
\end{figure*}

Representatively, we show in Fig.~\ref{fig:KS_bands} the paramagnetic Kohn-Sham (KS) band structure of NdN and TbN. The partially filled $Ln$ 4$f$ bands are metallic and located within the gap  between the N 2$p$ and $Ln$ 5$d$ bands. Due to the presence of Ln 4$f$ in-gap states, the $pd$ gap is enhanced to around \unit[3]{eV}. Along the $Ln$N series, the occupied $Ln$ 4$f$ states are moving towards higher binding energies. Hence, in the KS band structure of NdN they are mainly hybridizing with the conduction $Ln$ 5$d$ bands, while in TbN the 4$f$ bands mainly mix with the N 2$p$ valence bands. 

\vspace{4em}
\bibliographystyle{h-physrev}
\bibliography{nitrides}

\begin{thebibliography}{10}

\bibitem{Trodahl2007_GdN}
H.~J. Trodahl {\em et~al.},
\newblock Phys. Rev. B {\bfseries 76}, 085211 (2007).

\bibitem{Meyer2008}
C.~Meyer {\em et~al.},
\newblock Phys. Rev. B {\bfseries 78}, 174406 (2008).

\bibitem{Lambrecht2013}
F.~Natali {\em et~al.},
\newblock Progress in Materials Science {\bfseries 58}, 1316 (2013).

\bibitem{Hewett2019_NdN}
W.~F. Holmes-Hewett, R.~G. Buckley, B.~J. Ruck, F.~Natali, and H.~J. Trodahl,
\newblock Phys. Rev. B {\bfseries 100}, 195119 (2019).

\bibitem{Hewett2020_DyN}
W.~F. Holmes-Hewett {\em et~al.},
\newblock Applied Physics Letters {\bfseries 117}, 222409 (2020).

\bibitem{Duan2007}
C.-G. Duan {\em et~al.},
\newblock  {\bfseries 19}, 315220 (2007).

\bibitem{Senapati2011}
K.~Senapati, M.~G. Blamire, and Z.~H. Barber,
\newblock Nature Materials {\bfseries 10}, 849 (2011).

\bibitem{Caruso2019}
R.~Caruso {\em et~al.},
\newblock Phys. Rev. Lett. {\bfseries 122}, 047002 (2019).

\bibitem{Hewett2019_SmN}
W.~F. Holmes-Hewett, R.~G. Buckley, B.~J. Ruck, F.~Natali, and H.~J. Trodahl,
\newblock Phys. Rev. B {\bfseries 99}, 205131 (2019).

\bibitem{Azeem2018_SmN}
M.~Azeem,
\newblock Chinese Journal of Physics {\bfseries 56}, 1925 (2018).

\bibitem{Azeem2019_SmN}
M.~Azeem,
\newblock Advances in Natural Sciences: Nanoscience and Nanotechnology
  {\bfseries 10}, 015003 (2019).

\bibitem{Punya2011}
A.~Punya, T.~Cheiwchanchamnangij, A.~Thiess, and W.~R.~L. Lambrecht,
\newblock MRS Online Proceedings Library {\bfseries 1290}, 404 (2011).

\bibitem{Wachter1980}
P.~Wachter and E.~Kaldis,
\newblock Solid State Communications {\bfseries 34}, 241 (1980).

\bibitem{Degiorgi1990}
L.~Degiorgi, W.~Bacsa, and P.~Wachter,
\newblock Phys. Rev. B {\bfseries 42}, 530 (1990).

\bibitem{Wachter1998}
P.~Wachter {\em et~al.},
\newblock Solid State Communications {\bfseries 105}, 675 (1998).

\bibitem{Granville2006}
S.~Granville {\em et~al.},
\newblock Phys. Rev. B {\bfseries 73}, 235335 (2006).

\bibitem{Trodahl2016}
E.-M. Anton {\em et~al.},
\newblock Phys. Rev. B {\bfseries 93}, 064431 (2016).

\bibitem{Preston2007}
A.~R.~H. Preston {\em et~al.},
\newblock Phys. Rev. B {\bfseries 76}, 245120 (2007).

\bibitem{Azeem2013_DyN}
M.~Azeem {\em et~al.},
\newblock Journal of Applied Physics {\bfseries 113}, 203509 (2013).

\bibitem{Hasegawa1977}
A.~Hasegawa and A.~Yanase,
\newblock Journal of the Physical Society of Japan {\bfseries 42}, 492 (1977).

\bibitem{Aerts2004}
C.~M. Aerts {\em et~al.},
\newblock Phys. Rev. B {\bfseries 69}, 045115 (2004).

\bibitem{Larson2007}
P.~Larson, W.~R.~L. Lambrecht, A.~Chantis, and M.~van Schilfgaarde,
\newblock Phys. Rev. B {\bfseries 75}, 045114 (2007).

\bibitem{Morari2015}
C.~Morari {\em et~al.},
\newblock Journal of Physics: Condensed Matter {\bfseries 27}, 115503 (2015).

\bibitem{Doll2008}
K.~Doll,
\newblock Journal of Physics: Condensed Matter {\bfseries 20}, 075214 (2008).

\bibitem{Schlipf2011}
M.~Schlipf, M.~Betzinger, C.~Friedrich, M.~Le\ifmmode \check{z}\else
  \v{z}\fi{}ai\ifmmode~\acute{c}\else \'{c}\fi{}, and S.~Bl\"ugel,
\newblock Phys. Rev. B {\bfseries 84}, 125142 (2011).

\bibitem{Chantis2007}
A.~N. Chantis, M.~van Schilfgaarde, and T.~Kotani,
\newblock Phys. Rev. B {\bfseries 76}, 165126 (2007).

\bibitem{Lambrecht2015}
T.~Cheiwchanchamnangij and W.~R.~L. Lambrecht,
\newblock Phys. Rev. B {\bfseries 92}, 035134 (2015).

\bibitem{Metzner1989}
W.~Metzner and D.~Vollhardt,
\newblock Phys. Rev. Lett. {\bfseries 62}, 324 (1989).

\bibitem{Georges1992}
A.~Georges and G.~Kotliar,
\newblock Phys. Rev. B {\bfseries 45}, 6479 (1992).

\bibitem{Anisimov1997_1}
V.~I. Anisimov, A.~I. Poteryaev, M.~A. Korotin, A.~O. Anokhin, and G.~Kotliar,
\newblock Journal of Physics: Condensed Matter {\bfseries 9}, 7359 (1997).

\bibitem{Kotliar2006}
G.~Kotliar {\em et~al.},
\newblock Rev. Mod. Phys. {\bfseries 78}, 865 (2006).

\bibitem{Held2007}
K.~Held,
\newblock Advances in Physics {\bfseries 56}, 829 (2007).

\bibitem{Locht2016}
I.~L.~M. Locht {\em et~al.},
\newblock Phys. Rev. B {\bfseries 94}, 085137 (2016).

\bibitem{Pourovskii2009}
L.~V. Pourovskii, K.~T. Delaney, C.~G. Van~de Walle, N.~A. Spaldin, and
  A.~Georges,
\newblock Phys. Rev. Lett. {\bfseries 102}, 096401 (2009).

\bibitem{Peters2014}
L.~Peters {\em et~al.},
\newblock Phys. Rev. B {\bfseries 89}, 205109 (2014).

\bibitem{Lebeque2005}
S.~Leb\`egue {\em et~al.},
\newblock Phys. Rev. B {\bfseries 72}, 245102 (2005).

\bibitem{hubbard_1}
J.~Hubbard,
\newblock Proc. Roy. Soc. (London) {\bfseries A 276}, 238 (1963).

\bibitem{tran_mbj_original}
F.~Tran and P.~Blaha,
\newblock Phys. Rev. Lett. {\bfseries 102}, 226401 (2009).

\bibitem{koller_mbj_2011}
D.~Koller, F.~Tran, and P.~Blaha,
\newblock Phys. Rev. B {\bfseries 83}, 195134 (2011).

\bibitem{Galler2021_LnSF}
A.~Galler, J.~Boust, A.~Demourgues, S.~Biermann, and L.~V. Pourovskii,
\newblock Phys. Rev. B {\bfseries 103}, L241105 (2021).

\bibitem{Boust2021}
J.~Boust, A.~Galler, S.~Biermann, and L.~V. Pourovskii,
\newblock Combining semi-local exchange with dynamical mean-field theory:
  electronic structure and optical response of rare-earth sesquioxides,
  arXiv:2110.00400.

\bibitem{wien2k}
K.~Schwarz, P.~Blaha, and G.~Madsen,
\newblock Computer Physics Communications {\bfseries 147}, 71 (2002).

\bibitem{Aichhorn2009}
M.~Aichhorn {\em et~al.},
\newblock Phys. Rev. B {\bfseries 80}, 085101 (2009).

\bibitem{Aichhorn2011}
M.~Aichhorn, L.~Pourovskii, and A.~Georges,
\newblock Phys. Rev. B {\bfseries 84}, 054529 (2011).

\bibitem{triqs_cpc_2015}
O.~Parcollet {\em et~al.},
\newblock Computer Physics Communications {\bfseries 196}, 398  (2015).

\bibitem{triqs_dfttools}
M.~Aichhorn {\em et~al.},
\newblock Computer Physics Communications {\bfseries 204}, 200  (2016).

\bibitem{hong_mbj_2013}
H.~Jiang,
\newblock The Journal of Chemical Physics {\bfseries 138}, 134115 (2013).

\bibitem{Pourovskii2007}
L.~V. Pourovskii, B.~Amadon, S.~Biermann, and A.~Georges,
\newblock Phys. Rev. B {\bfseries 76}, 235101 (2007).

\bibitem{carnall_lanthanides_1989}
W.~T. Carnall, G.~L. Goodman, K.~Rajnak, and R.~S. Rana,
\newblock The Journal of Chemical Physics {\bfseries 90}, 3443 (1989).

\bibitem{Dederichs1984}
P.~H. Dederichs, S.~Bl\"ugel, R.~Zeller, and H.~Akai,
\newblock Phys. Rev. Lett. {\bfseries 53}, 2512 (1984).

\bibitem{Hybertsen1989}
M.~S. Hybertsen, M.~Schl\"uter, and N.~E. Christensen,
\newblock Phys. Rev. B {\bfseries 39}, 9028 (1989).

\bibitem{Gunnarsson1989}
O.~Gunnarsson, O.~K. Andersen, O.~Jepsen, and J.~Zaanen,
\newblock Phys. Rev. B {\bfseries 39}, 1708 (1989).

\bibitem{Anisimov1991}
V.~I. Anisimov and O.~Gunnarsson,
\newblock Phys. Rev. B {\bfseries 43}, 7570 (1991).

\bibitem{Cococcioni2005}
M.~Cococcioni and S.~de~Gironcoli,
\newblock Phys. Rev. B {\bfseries 71}, 035105 (2005).

\bibitem{Lichtenstein_LDApp}
A.~I. Lichtenstein and M.~I. Katsnelson,
\newblock Phys. Rev. B {\bfseries 57}, 6884 (1998).

\bibitem{Brown2012}
J.~D. Brown {\em et~al.},
\newblock Applied Physics Letters {\bfseries 100}, 072108 (2012).

\bibitem{Leuenberger2005}
F.~Leuenberger, A.~Parge, W.~Felsch, K.~Fauth, and M.~Hessler,
\newblock Phys. Rev. B {\bfseries 72}, 014427 (2005).

\bibitem{Preston2010_GdN}
A.~R.~H. Preston {\em et~al.},
\newblock Applied Physics Letters {\bfseries 96}, 032101 (2010).

\bibitem{Yamamoto2004}
T.~A. Yamamoto, T.~Nakagawa, K.~Sako, T.~Arakawa, and H.~Nitani,
\newblock Journal of Alloys and Compounds {\bfseries 376}, 17 (2004).

\end{thebibliography}

\end{document}